\documentclass[aps,twocolumn]{revtex4}
\usepackage{epsf}
\newcommand{\mb}[1]{ { \mbox{\boldmath{$#1$}}}  } 
 
\begin{document}
    
\title{Fano-type interference in quantum dots coupled 
       between metallic and superconducting leads}

\author{J.\ Bara\'nski  and  T.\ Doma\'nski}
\affiliation{
       Institute of Physics, M.\ Curie-Sk\l odowska University, 
       20-031 Lublin, Poland}

\date{\today}

\begin{abstract}
We analyze the quantum interference effects appearing in the
charge current through the double quantum dots coupled in 
$T$-shape configuration to an isotropic superconductor and 
metallic lead. Owing to proximity effect the quantum dots 
inherit a pairing which has the profound influence on  
nonequilibrium charge transport, especially in the subgap 
regime $|V| < \Delta/|e|$. We discuss under what conditions  
the Fano-type lineshapes might appear in such Andreev 
conductance and consider a possible interplay with the
strong correlation effects.
\end{abstract}

\maketitle

\section{Introduction}
Heterostructures with nanoobjects (such as quantum dots, 
nanowires, molecules, etc) hybridized to one conducting and 
another superconducting electrode seem to be promising testing 
fields where the strong electron correlations (responsible 
e.g.\ for Coulomb blockade and Kondo physics \cite{Pustilnik-04}) 
can be confronted with the superconducting order \cite{review-paper}. 
Coulomb repulsion between electrons in the solid state physics 
is known to suppress the local ($s$-wave) pairing and, through 
the spin exchange mechanism, eventually promotes the intersite 
($d$-wave) superconductivity \cite{plain-vanilla}. Mutual 
relation between such repulsion and the local pairing is 
however rather difficult for studying, both on theoretical 
grounds and experimentally. In nanoscopic heterostructures 
some of these limitations can be overcome by a suitable 
adjustment of the hybridization and the gate-voltage positioning 
of energy levels involved in the charge transfer \cite{book}. 
They enable a controllable changeover between 
the Kondo regime and opposite case dominated by the induced 
on-dot pairing.

Quantum dot (QD) coupled with the strength $\Gamma_{N}$ to metallic 
conductor (N) and with $\Gamma_S$ to superconducting electrode (S) 
can exhibit the features characteristic both for the on-dot pairing 
and the Kondo effect (including their coexistence) \cite{Deacon_etal}. 
Their efficiency depends on the ratio $\Gamma_{S}/\Gamma_{N}$. 
In the limit $\Gamma_{S} \gg \Gamma_{N}$ the underlying physics 
is controlled by on-dot pairing and manifests itself e.g.\ 
by the particle-hole splitting of the quasiparticle levels. 
On the other hand for $\Gamma_{S} \ll \Gamma_{N}$ the strong 
correlations take over. Non-trivial aspects related to such interplay 
between the Coulomb interactions and the proximity induced on-dot 
pairing has been addressed theoretically using various methods 
like: the mean field slave boson approach \cite{Fazio-98}, 
the noncrossing approximation \cite{Clerk-00}, perturbative 
scheme \cite{Cuevas-01,review-paper}, constrained slave boson 
technique \cite{Krawiec-04}, numerical renormalization group 
\cite{Tanaka-07,Bauer-07,Hecht-08} and other 
\cite{Sun-99,Cho-99,Avishai-01,Domanski-08}.
Also the cotunneling regime of a Coulomb blockaded quantum dot sandwiched
between a normal and superconducting lead, where charge fluctuations
are strongly suppressed, has been discussed emphasizing 
the role of in-gap resonances \cite{Paaske-10}.

As far as the experimental situation is concerned it has been less 
intensively explored. The earliest transport measurements for N-QD-S 
interface have been obtained using the multi-wall carbon nanotube 
deposited between Au and Al electrodes \cite{Graeber-04}. Those
investigations concentrated however on the specific regime $k_{B}
T_{K} \geq \Delta$, when the Coulomb correlations dominated over 
the proximity effect. Other studies of the same group have been done 
for similar structures replacing a metallic electrode by a ferromagnet 
\cite{Hofstetter-10}. Several recent efforts focused on the 
multiterminal structures involving two normal and one superconducting 
electrodes as useful schemes for realization of: the crossed 
Andreev reflections tunable via gate voltages \cite{Herrmann-10},
the Cooper pair splitters \cite{Eldridge-10}, and the QD spin 
valves \cite{Sothmann-10}.

\begin{figure}
\epsfxsize=6.5cm\centerline{\epsffile{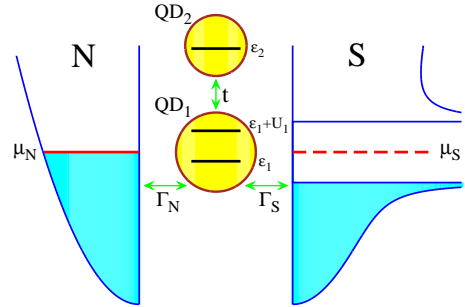}}
\caption{(color online) Scheme of the double quantum dot 
in $T$-shape configuration coupled to the conducting (N) and 
superconducting (S) leads, where interference effects
originate from the interdot hopping $t$.}
\label{scheme}
\end{figure}

Very useful understanding of a subtle interplay between 
the correlations and the induced on-dot pairing has been gained 
from recent measurements by R.S.\ Deacon {\em et al} \cite{Deacon_etal}.
The authors have explored the subgap transport originating 
from the Andreev-type scattering processes for several 
representative ratios $\Gamma_{S}/\Gamma_{N}$ using the 
self-assembled InAs quantum dots deposited between the golden 
(N) and aluminium (S) electrodes. Their measurements provided
the unambiguous experimental evidence for: 
a) particle-hole splitting of the subgap conductance of   
the Andreev states when $\Gamma_{S} \geq \Gamma_{N}$, and  
b) enhancement of the zero-bias Andreev conductance due to 
formation of the Kondo resonance at the Fermi level of metallic 
lead, as has been qualitatively suggested by our studies
\cite{Domanski-08} and also indicated by other groups 
\cite{Yamada-10}. 

The present work extends our former studies by taking into account 
interference effects arising from the additional degrees of 
freedom. As the simplest prototype for Fano-type interference 
\cite{Miroshnichenko-10} we consider the setup (see figure 
\ref{scheme}) with a side-attached quantum dot contributing
 an extra pathway for electrons transmitted between the 
metallic and superconducting leads. Our analysis 
is complementary to the previous study by Y.\ Tanaka 
{\em et al} who considered the double quantum dots coupled
between N and S leads in a $T$-shape setup but assuming 
$U_{1}\!=\!0$, $U_{2}\!\neq\!0$ \cite{Tanaka-08} and in
a series configuration \cite{Tanaka-10}.

In section 2 we introduce the microscopic model and briefly 
outline basic notes on the nonequilibrium subgap transport. 
In the next section 3 we discuss a unique way in which the 
Fano-type lineshapes might be observed in Andreev conductance, 
focusing on the uncorrelated quantum dots. In the last part 
(section 4) we discuss the influence of correlations at 
the interfacial quantum dot which seem to have remarkable 
influence on the low bias transport. We end with the summary 
and suggestions for the future studies.

\section{The model}

For description of the heterojunction illustrated 
in figure \ref{scheme} we use the  Hamiltonian
\begin{eqnarray} 
\hat{H} = \hat{H}_{N} + \hat{H}_{N-DQD} +  \hat{H}_{DQD}
        + \hat{H}_{S-DQD} + \hat{H}_{S} 
\label{model} 
\end{eqnarray}
where the double quantum dot (DQD) is represented by
\begin{eqnarray} 
 \hat{H}_{DQD} \!=\! \sum_{\sigma  i} \epsilon_{i} 
\hat{d}^{\dagger}_{i \sigma} \hat{d}_{i \sigma}  
\!+\!  U_{1} \; \hat{n}_{1 \uparrow} \hat{n}_{1 \downarrow}  
+ \left( t \; \hat{d}_{1\sigma}^{\dagger}  
\hat{d}_{2\sigma} \!+\! \mbox{\rm h.c.} \right) \!.
\label{DQD} 
\end{eqnarray} 
The energies of each ($i=1,2$) quantum dot electrons are denoted 
by $\varepsilon_{i}$ and $t$ stands for the usual interdot hopping. 
We restrict considerations of the correlation effects (section 4) 
to the Coulomb repulsion $U_{1}$ between opposite spin electrons 
$\sigma=\uparrow$, $\downarrow$ at the interfacial quantum dot.

The external reservoirs N and S of charge carriers are described 
by $\hat{H}_{N} \!=\! \sum_{{\bf k},\sigma} \xi_{{\bf k}N}  
\hat{c}_{{\bf k} \sigma N}^{\dagger} \hat{c}_{{\bf k} \sigma N}$ 
and correspondingly $\hat{H}_{S} \!=\!\sum_{{\bf k},\sigma}  \xi_{{\bf k}S}
\hat{c}_{{\bf k} \sigma S }^{\dagger}  \hat{c}_{{\bf k} \sigma S} 
\!-\! \sum_{\bf k} \Delta   \hat{c}_{{\bf k} \uparrow S }^{\dagger} 
\hat{c}_{-{\bf k} \downarrow S }^{\dagger} \!+\! \Delta^{*} 
\hat{c}_{-{\bf k} \downarrow S} \hat{c}_{{\bf k} \uparrow S}$  
assuming the isotropic energy gap $\Delta$. As usually 
$\xi_{{\bf k}\beta}\!=\!\varepsilon_{{\bf k}\beta} \!-\!\mu_{\beta}$ 
denote the electron energies measured from the individual chemical 
potentials $\mu_{\beta}$ which become detuned $\mu_{N}\!-\!\mu_{S}
\!=\!eV$ if a bias $V$ is applied across the junction inducing the 
nonequilibrium charge flow $I(V)$. Fano-type quantum interference 
effects originating from the hopping $t$ to side-coupled quantum 
dot $i\!=\!2$ are discussed here assuming that only the 
interfacial quantum dot  $i\!=\!1$ is directly coupled to 
external leads  
\begin{eqnarray} 
\hat{H}_{\beta - DQD} = \sum_{{\bf k},\sigma } \left(
 V_{{\bf k} \beta} \; \hat{d}_{1 \sigma}^{\dagger}  
\hat{c}_{{\bf k} \sigma \beta } + \mbox{\rm h.c.} \right) .
\label{hybr}
\end{eqnarray} 
In the wide-band limit approximation it is convenient to introduce 
the structureless coupling constants 
$\Gamma_{\beta}=2\pi\sum|V_{{\bf k} \beta}|^{2}
\delta\left( \omega\!-\!\xi_{\bf k}\right)$
which shall be used here as the energy units. 

\begin{figure}
\epsfxsize=8cm\centerline{\epsffile{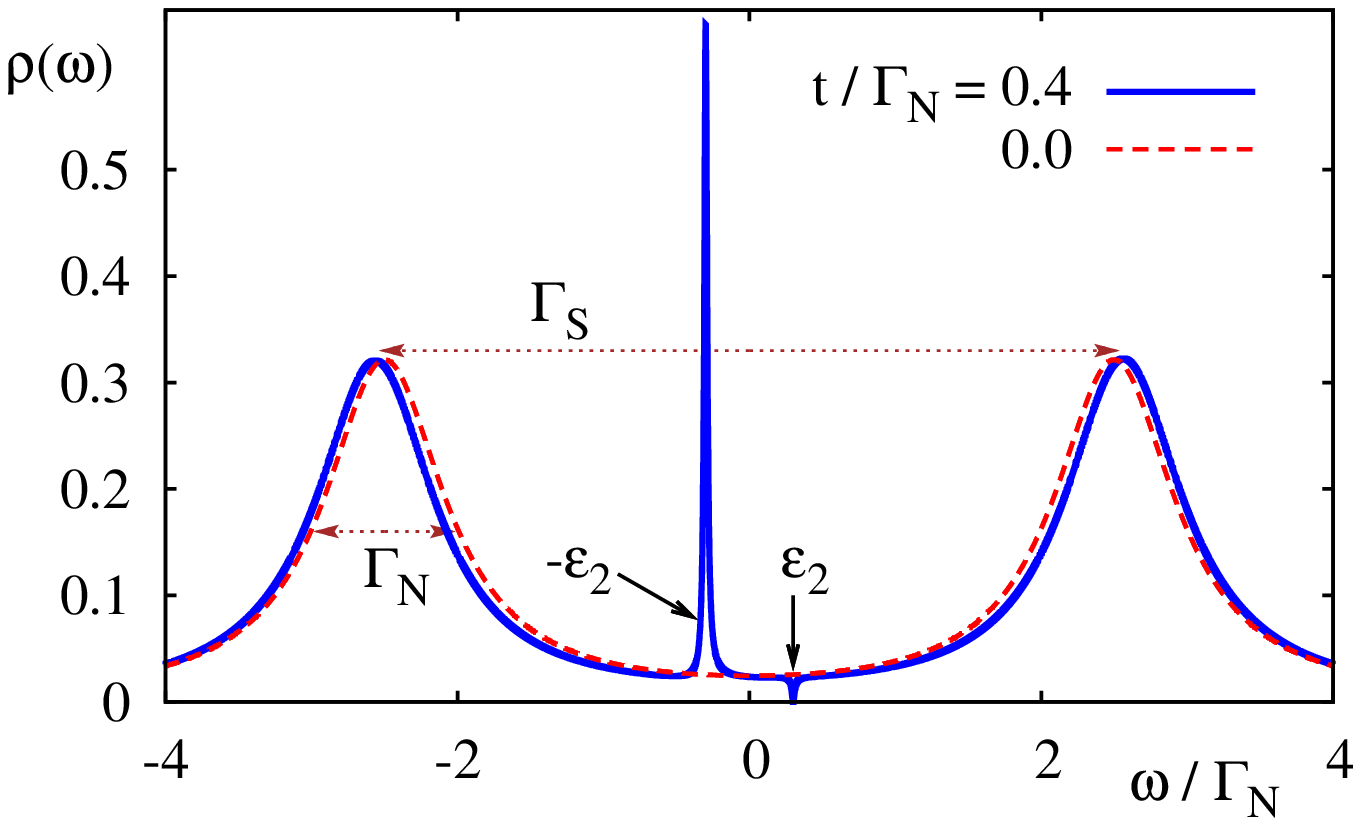}}
\epsfxsize=8cm\centerline{\epsffile{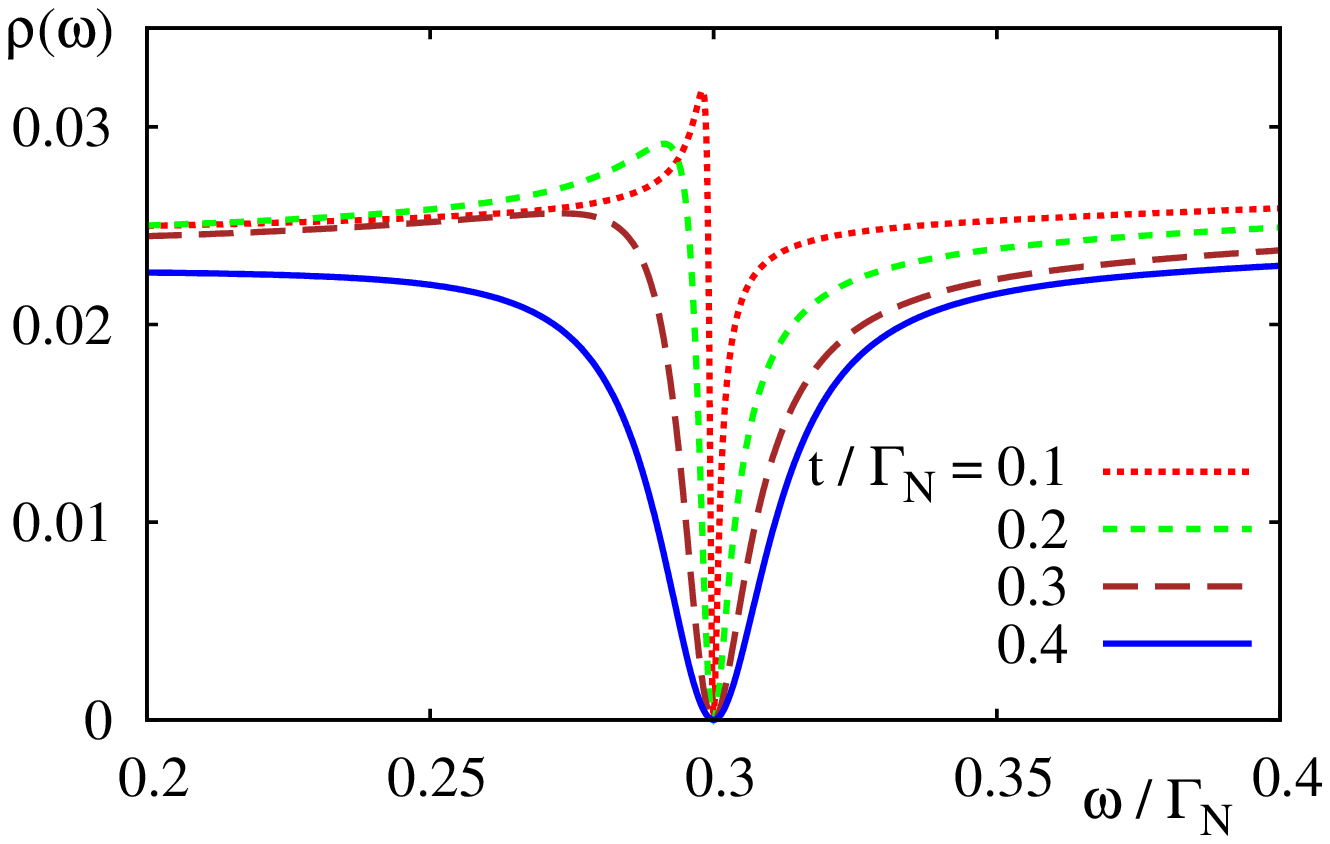}}
\caption{(color online) Density of states $\rho(\omega)$ 
of the interfacial $i\!=\!1$ QD obtained in equilibrium 
conditions for $\varepsilon_{1}=0$, $\varepsilon_{2}
\!=\!0.3\Gamma_{N}$, $\Gamma_{S}\!=\!5 \Gamma_{N}$ and 
a few values of the interdot hopping $t$.}
\label{dos}
\end{figure}

Interplay between the proximity induced on-dot pairing, the 
correlations and the quantum interference effects can be in 
practice detected by measuring the differential conductance 
$dI(V)/dV$. Particularly valuable for this purpose is the 
low voltage (subgap) regime $|eV| \ll \Delta$. Under such 
conditions the charge current is provided by the anomalous 
Andreev scattering in which electrons from the metallic lead 
are converted into the Cooper pairs in superconductor with
a simultaneous reflection of the electron holes back 
to the normal lead. On a formal level the resulting Andreev 
current can be expressed by the Landauer-type 
formula \cite{Krawiec-04,Sun-99}
\begin{eqnarray} 
I_{A}(V) = \frac{2e}{h} \int d\omega T_{A}(\omega)
\left[ f(\omega\!-\!eV,T)-f(\omega\!+\!eV,T)\right],
\label{I_A}
\end{eqnarray} 
where $f(\omega,T)$ is the Fermi distribution function 
and the transmittance $T_{A}(\omega)=\Gamma_{N}^{2} 
\left| G_{12}(\omega) \right|^{2}$ depends on the 
off-diagonal part (in the Nambu notation) of the 
retarded Green's function (\ref{GF}) of the interfacial 
quantum dot.

\section{Fano resonances}

Fano resonances appear in many physical systems due to the 
quantum interference of the waves transmitted resonantly via some 
discrete energy level combined with transmittance contributed 
from a continuum of other states. In nanoscale physics such 
resonances are feasible in a variety of constructions 
\cite{Miroshnichenko-10}. Fano lineshapes are present for instance
in the electron transport when two external electrodes are in 
parallel coupled through a quantum dot and directly via a shortcut 
bridge \cite{Bulka-01}. Another simple possibility takes place in 
the electron tunneling using two quantum dots with considerably 
different linebroadenings \cite{Zitko-10,Trocha-07}. In the 
latter case the narrower level is responsible for forming 
the Fano resonance on a background of the broader level.

In this work we want to analyze similar interference effects 
appearing in the anomalous Andreev current, which is very specific 
because of the particle and hole degrees of freedom mixed with one 
and other. To have a clear picture of the underlying physics let 
us start by considering the noninteracting case $U_{i}\!=\!0$ when 
the Green's functions of each quantum dot can be determined exactly. 

\begin{figure}
\epsfxsize=9cm\centerline{\epsffile{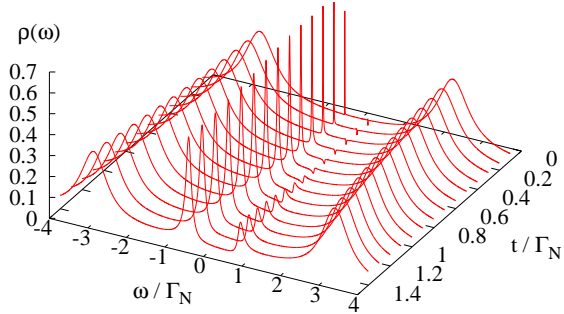}}
\caption{(color online) Changeover of the interfacial quantum 
dot spectrum from the Fano (resonance and antiresonance) 
lineshapes to the effective four-peak structure upon increasing
the interdot hopping $t$ for the same parameters 
as in figure \ref{dos}.}
\label{dos_evol}
\end{figure}

Electron transport of the setup shown in figure \ref{scheme} is 
determined by effective properties of the interfacial quantum dot. 
For this purpose we compute the matrix Green's function ${\mb G}
(\tau_{1},\tau_{2})\!=\!-\hat{T}_{\tau}\langle \hat{\Psi}(\tau_{1}) 
\hat{\Psi}^{\dagger}(\tau_{2})\rangle$ introducing the standard 
spinor notation $\hat{\Psi}^{\dagger}\!=\!(\hat{d}_{1\uparrow}
^{\dagger},\hat{d}_{1\downarrow})$ and $\hat{\Psi}\!=\!(\hat
{\Psi}^{\dagger})^{\dagger}$. In the equilibrium conditions 
$\mu_{N}\!=\!\mu_{S}$ this function depends only on the time 
difference and its Fourier transform obeys the following Dyson 
equation
\begin{eqnarray} 
{\mb G}(\omega)^{-1} = 
\left( \begin{array}{cc}  
\omega\!-\!\varepsilon_{1} &  0 \\ 0 &  
\omega\!+\!\varepsilon_{1}\end{array}\right)
- {\mb \Sigma}^{0}(\omega)  
- {\mb \Sigma}^{U}(\omega)  ,
\label{GF}\end{eqnarray} 
where the term ${\mb \Sigma}^{0}(\omega)$ corresponds to 
the selfenergy of noninteracting case ($U_{1}\!=\!0$) and  
${\mb  \Sigma}^{U}(\omega)$ accounts for the correlation 
effects (discussed in section IV). Focusing on the deep
subgap regime $|\omega| \ll \Delta$ we obtain that 
${\mb \Sigma}^{0}(\omega)$ simplifies to (see the appendix)
\begin{eqnarray} 
{\mb \Sigma}^{0}(\omega) = 
\left( \begin{array}{cc}  - \; \frac{i \Gamma_{N}}{2}  
+ \frac{t^{2}}{\omega-\varepsilon_{2}}
& - \; \frac{ \Gamma_{S}}{2} \\  
- \; \frac{ \Gamma_{S}}{2} 
& - \; \frac{i \Gamma_{N}}{2}  + \frac{t^{2}}{\omega+\varepsilon_{2}}
\end{array} \right)  .
\label{Sigma0}
\end{eqnarray} 
When the interference effects caused by the hopping $t$ to the 
side-coupled QD are neglected the expression (\ref{Sigma0}) 
becomes static ($\omega$-independent) and nontrivial physics of this, 
so called {\em atomic superconducting limit}, has been explored in detail 
by several groups \cite{review-paper,Tanaka-07,Florens-09} 
including ourselves \cite{Domanski-08}.

Taking into account the quantum interference  $t\!\neq\!0$ we show 
in figure \ref{dos} the  proximity induced on-dot pairing [formally 
arising from the off-diagonal parts of (\ref{Sigma0})] illustrating 
the energy spectrum $\rho(\omega) \!=\!-\frac{1}{\pi}{\mbox{\rm Im}}
{\mb G}_{11} (\omega+i0^{+})$ obtained for strong coupling to 
the superconducting lead $\Gamma_{S}\!=\!5\Gamma_{N}$. Such 
coupling $\Gamma_{S}$ is responsible for the particle-hole 
splitting of the effective quasiparticle states formed at 
$\pm\sqrt{\varepsilon_{1}^{2}\!+\!(\Gamma_{S}/2)^{2}}$ 
whereas the coupling $\Gamma_{N}$ controls their broadening. 
In the particular case $\varepsilon_{1}\!=\!0$ the quasiparticle 
peaks appearing at $\pm E_{1}$ (where $E_{1}\!\equiv\!
\sqrt{\varepsilon_{1}^{2}\!+\!\Gamma_{S}^{2}/4}$) are 
symmetric, but for arbitrary $\varepsilon_{1}$ they are 
weighted by the corresponding BCS coefficients $u^{2},v^{2}
\!=\!\frac{1}{2}\left( 1 \pm \varepsilon_{1}/E_{1} \right)$  
\cite{Domanski-08}. On top of such behavior we clearly notice 
that hopping to the side-coupled quantum dot induces additional 
features appearing in the effective spectrum near $\pm 
\varepsilon_{2}$ as the Fano resonance and antiresonance. 
For the case of both metallic leads there would survive just 
the single Fano structure at $\varepsilon_{2}$ which in 
very pedagogical way has been discussed by R.\ \v{Z}itko 
\cite{Zitko-10}. 

\begin{figure}
\epsfxsize=7cm\centerline{\epsffile{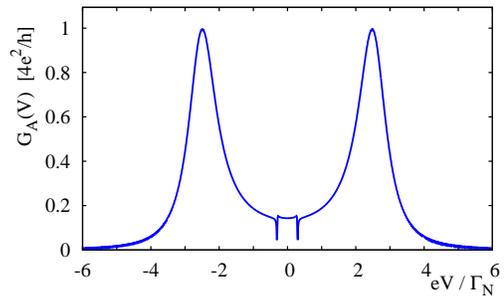}}
\caption{(color online) The differential Andreev conductance 
$G_{A}(V)$ versus the bias $V$ revealing the quasiparicle peaks 
(near $\pm\sqrt{\varepsilon_{1}^{2}\!+\!(\Gamma_{S}/2)^{2}}$) 
and Fano-type lineshapes (near $\pm \varepsilon_{2}$) for 
the set of parameters used in figure \ref{dos} and $t\!=\!
0.1\Gamma_{N}$.}
\label{GA}
\end{figure}

Fano-type lineshapes  (see the lower panel in figure \ref{dos}) 
are present only in the weak hopping regime $t \ll \Gamma_{N}$. 
For increasing $t$ the Fano structures gradually evolve into 
separate quasiparticle peaks illustrated in figure \ref{dos_evol}. 
Physically this can be assigned to the induced pairing on the 
side-attached QD $\langle \hat{d}_{2\downarrow} \hat{d}_
{2\uparrow}\rangle \neq 0$ transmitted there indirectly via 
the interfacial quantum dot. Such effect again qualitatively 
differs from the structures of the DQD coupled to both metallic 
leads \cite{Zitko-10,Trocha-07}. 

\begin{figure}
\epsfxsize=7cm\centerline{\epsffile{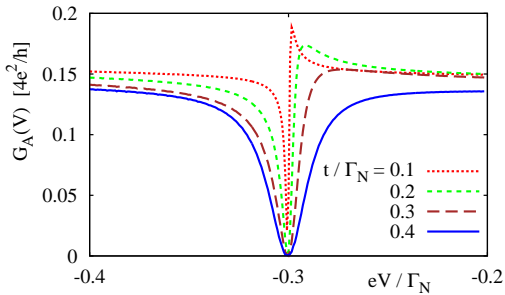}}
\caption{(color online) Differential conductance $G_{A}(V)$ 
of the subgap Andreev current versus the source-drain bias 
$V$ in a vicinity of the Fano structure appearing at 
$V=\pm \varepsilon_{2}/e$.}
\label{GA_fano}
\end{figure}
Interrelation between the interference and proximity effect 
can be practically investigated by measuring the tunneling 
current. In figure \ref{GA} we show bias voltage $V$ 
dependence of the differential Andreev conductance $G_{A}(V)
\!=\!dI_{A}(V)/dV$ determined at zero temperature from 
(\ref{I_A}) over a broad regime covering both the subgap 
quasiparticle peaks. Figure \ref{GA_fano} illustrates the 
resulting Fano-type lineshapes $G_{A}=G_{0} \frac{\left( 
x + q \right)^{2}}{x^{2}+1}+G_{1}$ nearby $-\varepsilon_{2}$, 
where $x\!=\!\left| eV\!+\!\varepsilon_{2}\right|/\Gamma_{N}$
and the asymmetry parameter $q$ gradually decreases upon 
increasing the hopping integral $t$. Our results can be thought as 
extension of the predictions  obtained for the normal electron 
tunneling using the $T$-shape DQD coupled to both metallic leads
\cite{Zitko-10,Sasaki-09} onto the anomalous Andreev current
where the particle hole mixing has the essential importance.

\vspace{-0.2cm}
\section{Interplay with correlations}

Coulomb repulsion between electrons of opposite spins can have 
an important influence on the spectral and transport properties 
of various nanostructures. For the case of quantum dots coupled 
to both conducting leads such interactions are known to be 
responsible for: a) the charging effect (if a given energy 
level $\varepsilon_{i}$ is attempted to be occupied by more 
than a single electron this costs the system an extra energy 
$U_{i}$), b) the Kondo effect when the singlet state is formed 
between QD and itinerant electrons from the leads  \cite{EOM}. 
In spectroscopic properties they are manifested by appearance of 
the Coulomb satellite around $\omega\!=\!\varepsilon_{i}
\!+\!U_{i}$ and the narrow Kondo resonance at the Fermi 
level. For heterostructures with the superconducting 
electrodes the situation is more complex due to 
a competition between the induced on-dot pairing 
and Coulomb repulsion.

The rich interplay between the quantum interference, correlations 
and proximity effect for the configuration shown in figure 
\ref{scheme} have been so far addressed using the density 
functional technique  \cite{Karmanyos-09} (which does not 
capture the Kondo physics) and by the numerical 
renormalization group calculations \cite{Tanaka-08}. 
In latter case the authors focused on $U_{1}\!=\!0$, 
$U_{2}\!\neq\!0$ when the side-attached quantum dot 
can indirectly form the Kondo state with electrons 
of the metallic lead  affecting the Andreev transport.

\begin{figure}
\epsfxsize=8cm\centerline{\epsffile{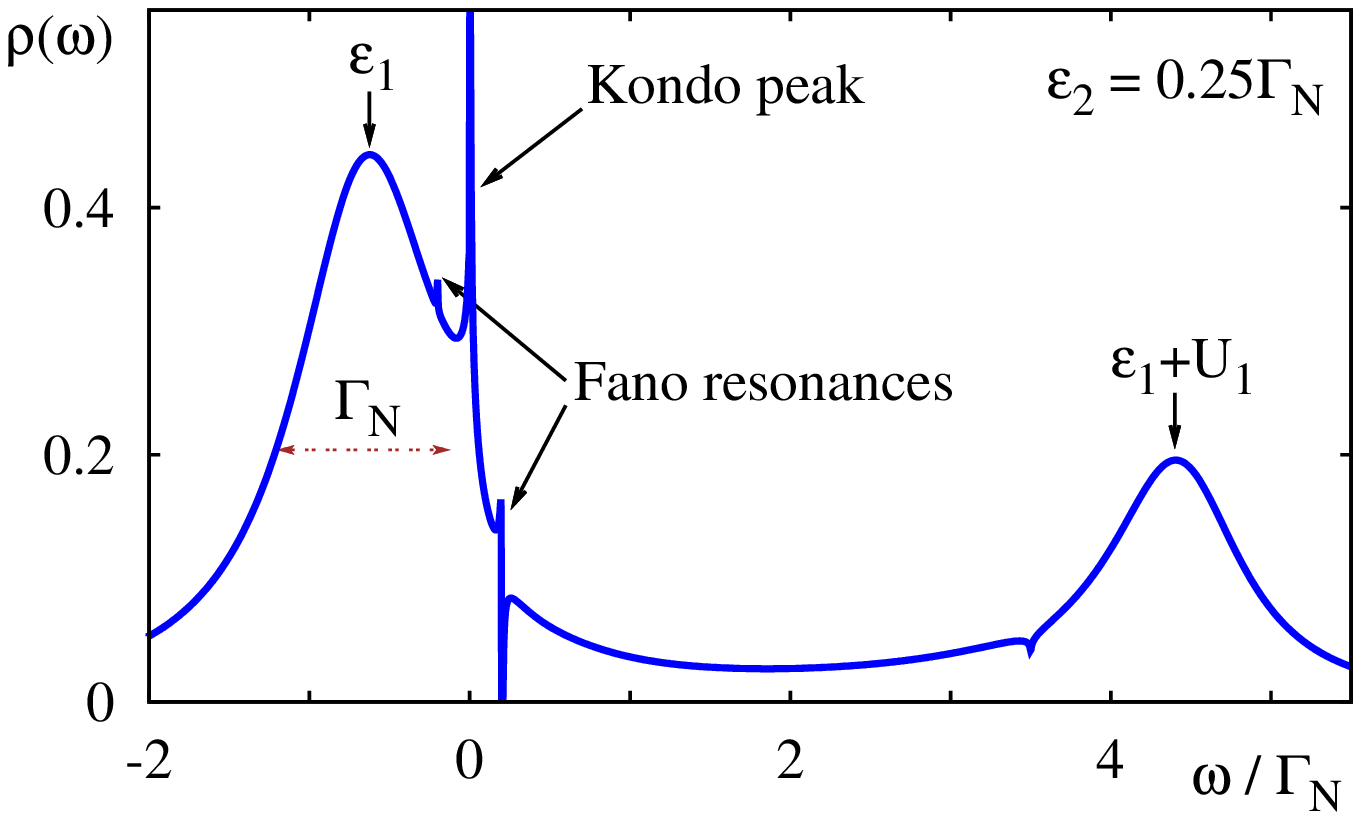}}
\epsfxsize=8cm\centerline{\epsffile{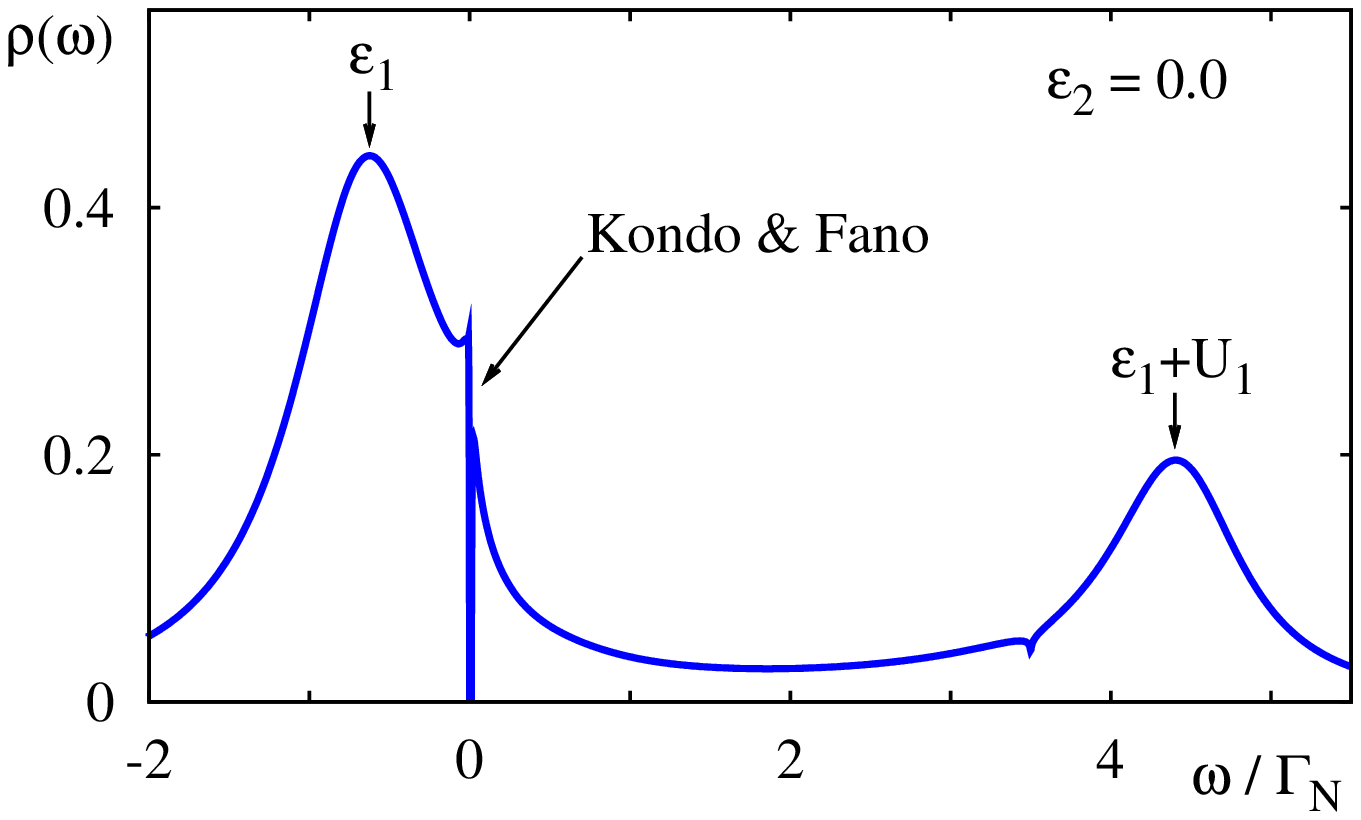}}
\caption{(color online) Density of states $\rho(\omega)$ 
of the correlated interfacial QD in the Kondo regime 
obtained for: $\varepsilon_{1}\!=\!-0.75\Gamma_{N}$,
$\Gamma_{S}\!=\!\Gamma_{N}$, $U\!=\!5\Gamma_{N}$, 
$t\!=\!0.1\Gamma_{N}$ and temperature $k_{B}T=0.001
\Gamma_{N}$. The upper panel shows the spectrum for 
$\varepsilon_{2}\!=\!0.25\Gamma_{N}$ with two Fano 
structures appearing at $\pm\varepsilon_{2}$ aside 
the Kondo peak. The lower plot corresponds to 
$\varepsilon_{2}\!=\!0$ when both the Kondo and 
Fano structures are superimposed.}
\label{corr_dos}
\end{figure}
 
To account for the correlation effects predominantly 
originating from the interfacial quantum dot we extend 
here the procedure previously used by us for studying 
the single quantum dot \cite{Domanski-08}. The main idea 
is to approximate the correlation selfenergy 
${\mb \Sigma}^{U}(\omega)$ by the diagonal matrix  
\begin{eqnarray} 
{\mb \Sigma}^{U}(\omega) \simeq 
\left( \begin{array}{cc}  \Sigma_{N}(\omega)
& 0 \\  0 
& - \Sigma_{N}^{*}(-\omega) 
\end{array} \right)  .
\label{U_correction}
\end{eqnarray} 
Such assumption (applied also in the NRG studies 
\cite{Tanaka-07}) can be thought as the simplest ansatz 
for the many-body selfenergy ${\mb \Sigma}^{U}(\omega)$ 
allowing to combine the proximity effect (\ref{Sigma0}) 
with the correlations, brought separately from the 
particle and hole channels. In more advanced treatments 
one should take into account the possible feedback effects 
between these normal and anomalous channels. We 
nevertheless hope that by imposing (\ref{U_correction}) 
we can get some insight at least on a qualitative level
which might stimulate the future studies. 

Within qualitative studies of the correlation effects 
we can describe the Coulomb blockade and Kondo effects 
using the following equation of motion expression \cite{EOM}
\begin{eqnarray}
\Sigma_{N}(\omega) =  
\omega \!-\! \varepsilon_{1}\!-\!
\frac{[\tilde{\omega}\!-\!\varepsilon_{1}]
[\tilde{\omega}\!-\!\varepsilon_{1}\! -\!U_{1}
\!-\!\Sigma_{3}(\omega)]\!+\!U_{1} \Sigma_{1}(\omega)} 
{\tilde{\omega}-\varepsilon_{1} - [
\Sigma_{3}(\omega)+U_{1}(1\!-\!n_{1,\sigma})]}
\label{ansatz}
\end{eqnarray}
where $\Sigma_{\nu\!=\!1,3}(\omega)\!=\!\sum_{\bf k} 
|V_{{\bf k} N}|^{2} \left[ f(\omega,T) \right]^
{\frac{3 - \nu }{2}}[(\omega\! - \! \xi_{{\bf k}N}
)^{-1} + (\omega\! -\! U_{1} \! - 2 \varepsilon_{1}
\!+\!\xi_{{\bf k} N})^{-1}]$, $n_{1,\sigma}\!=\!
\langle \hat{d}_{1\sigma}^{\dagger} \hat{d}_{1\sigma} \rangle$
 and $\tilde{\omega}
\!=\!\omega+\frac{i\Gamma_{N}}{2}$.
We explored the interfacial quantum dot spectrum and the 
related transport properties at $k_{B}T\!=\!0.001\Gamma_{N}$, 
i.e.\ well below the Kondo temperature. Specific numerical 
computations have been done for $\varepsilon_{1}\!=\!
-0.75\Gamma_{N}$, $U_{1}\!=\!5\Gamma_{N}$  and symmetric 
coupling to both external leads giving the optimal 
conditions for any low-bias features in the Andreev 
current \cite{Domanski-08,Deacon_etal}. This ratio 
$\Gamma_{S}/\Gamma_{N}\!\sim\!1$ is a reason why the 
particle-hole splitting is hardly visible, but otherwise
(for larger $\Gamma_{S}$) the Kondo peak is either 
reduced or completely absent \cite{Domanski-08}.

\begin{figure}
\epsfxsize=8cm\centerline{\epsffile{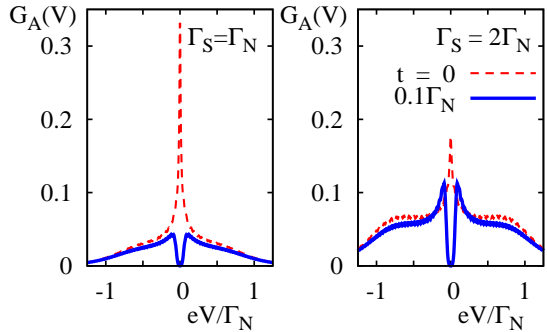}}
\caption{(color online) Andreev conductance $G_{A}(V)$ 
expressed in units of $4e^{2}/h$ obtained for $\Gamma_{S}
\!=\!\Gamma_{N}$ (left h.s.\ panel) and $\Gamma_{S}\!=\!
2\Gamma_{N}$ (panel on the right). The thin dashed lines 
correspond to $t\!=\!0$ whereas the thick solid lines show 
influence of the Fano-type interference for $t\!=\!0.1\Gamma_{N}$. 
In both plots we used $\varepsilon_{2}\!=\!0$ and the same
model parameters as in figure \ref{corr_dos}.}
\label{Kondo_peak}
\end{figure}

Upper panel in the figure \ref{corr_dos} illustrates the Fano 
resonance/antiresonance around $\pm\varepsilon_{2}$ (where 
$\varepsilon_{2}\!=\!0.25\Gamma_{N}$) obtained for the 
hopping $t\!=\!0.1\Gamma_{N}$. These Fano-type interference 
objects appear on top of the characteristic spectrum consisting 
of the Kondo resonance and the broad quasiparticle peaks 
seen at $\varepsilon_{1}$ and its Coulomb satellite at 
$\varepsilon_{1}\!+\!U_{1}$. Such spectrum is the result 
of interference effects discussed in the preceding section 
and the correlation features. The lower panel of figure 
\ref{corr_dos} shows the very specific situation 
$\varepsilon_{2}\!=\!0$ when the Kondo and Fano structures 
coincide with each other. The Fano-type resonance seems to 
play then a dominant role.

Differential conductance of the Andreev current (\ref{I_A})
for the $T$-shaped double quantum dot system (\ref{model})  
is shown in figure \ref{Kondo_peak}. We notice a clear
suppression of the zero-bias peak (present for the single
quantum dot case $t\!=\!0$ as indicated by the dashed
lines) because of a destructive influence of the 
Fano-type interference (the solid lines). 
The subgap Andreev current is thus very sensitive to 
the interplay between the Kondo and Fano effects. For 
obvious reasons their most dramatic competition occurs 
when energy of the side-coupled quantum dot coincides 
with the Kondo resonance, i.e.\ for $\varepsilon_{2}
\!=\!0$. Quantum interference effects destroy 
then the Kondo peak washing out the zero-bias 
enhancement of the Andreev conductance.  

\section{Summary}

We have studied a unique nature in which the Fano-type 
quantum interference  manifest itself in the energy spectrum 
and differential conductance of the heterojunction 
where a metallic lead is coupled via double quantum dot  
to superconducting electrode. In the regime of subgap 
source-drain voltage $|eV| < |\Delta|$ nonequilibrium 
charge transport is contributed only through the anomalous 
Andreev channel when electron  from the metallic electrode 
is converted into the Cooper pair (propagating further in 
superconductor) with a simultaneous reflection of hole 
back to the metallic lead. Transmittance of such Andreev
scattering is a sensitive probe of the proximity induced 
on-dot pairing as well as the quantum interference 
and correlation effects.

Since on-dot pairing mixes the particle with hole states 
the interference effects are doubled in a comparison to 
similar junctions  without the superconducting electrode. In 
particular, for $T$-shape configuration schematically shown in 
figure \ref{scheme} we notice that effective spectrum of the 
interfacial quantum dot develops the resonance and antiresonance, 
correspondingly at $\pm \varepsilon_{2}$ (figure \ref{dos}). 
These Fano-type structures are present whenever the hopping 
integral $t$ to the side-attached quantum dot ($i\!=\!2$) 
is much smaller than the linebroadening $\Gamma_{N}$ 
(whereas $\Gamma_{S}$ merely controls the induced 
quasiparticle splitting). Upon increasing $t$ the Fano-type 
features disappear, evolving into the new quasiparticle 
peaks (figure \ref{dos_evol}) being a consequence of 
the proximity effect indirectly spread onto 
the side-attached quantum dot. 

Correlation effects play an important role with regard to 
the following aspects: a) the charging effect which causes 
appearance of the Coulomb satellite near $\varepsilon_{1}\!+\!U_{1}$,
b) the Kondo singlet state (when the interfacial quantum dot 
spin is effectively screened by electrons of the metallic 
electrode leading to formation of a narrow resonance at 
$\mu_{N}$), and c) eventual suppression the on-dot pairing.
We have previously shown \cite{Domanski-08} that the Kondo 
effect enhances the zero-bias Andreev conductance as  
indeed reported experimentally \cite{Deacon_etal}. 
In the present work we indicate that in the double
quantum dots the quantum interference can (destructively) 
affect such feature if the Fano-type structures appear 
nearby the Kondo peak. 

A more detailed analysis of the Fano-Kondo interplay 
could be a challenging task in the future studies. 
For this purpose one should resort either to nonperturbative 
techniques, like the numerical renormalization group, or 
to some sophisticated perturbative methods capable to  
interpolate between the limits $t \rightarrow 0$,
$\Gamma_{\beta} \rightarrow 0$ and $U \rightarrow 0$.

\begin{acknowledgments}
We thank B.\ Bu\l ka and K.I.\ Wysoki\'nski for instructive 
discussions on the Fano resonances in nanophysics.
This work is partly supported by the Polish Ministry 
of Science and Education under the grant NN202187833.
\end{acknowledgments}

\appendix*
\section{Selfenergy of the noncorrelated DQD}

\noindent
Using the standard Nambu notation we can express the 
retarded Green's functions of the metallic lead
\begin{eqnarray} 
g_{N}^{r}({\bf k}, \omega) = 
\left( \begin{array}{cc}  
\frac{1}{\omega-\xi_{{\bf k}N}} & 0 \\ 
0 &  
\frac{1}{\omega+\xi_{{\bf k}N}}
\end{array}\right)
\label{gN}
\end{eqnarray} 
the (unperturbed) side-attached quantum dot
\begin{eqnarray} 
g_{2}^{r}(\omega) = \left( \begin{array}{cc}  
\frac{1}{\omega-\varepsilon_{2}} & 0 \\ 
0 &  \frac{1}{\omega+\varepsilon_{2}}
\end{array} \right)
\label{g_2}
\end{eqnarray} 
and the isotropic superconductor 
\begin{eqnarray} 
g_{S}^{r}({\bf k}, \omega) = 
\left( \begin{array}{cc}  
\frac{u^{2}_{\bf k}}{\omega-E_{\bf k}}+\frac{v^{2}_{\bf k}}
{\omega+E_{\bf k}} \hspace{0.2cm} & \frac{-u_{\bf k}v_{\bf k}}
{\omega-E_{\bf k}}+\frac{u_{\bf k}v_{\bf k}}{\omega+E_{\bf k}}
\\ 
\frac{-u_{\bf k}v_{\bf k}}{\omega-E_{\bf k}}+
\frac{u_{\bf k}v_{\bf k}}{\omega+E_{\bf k}}
& \frac{u^{2}_{\bf k}}{\omega+E_{\bf k}}+
\frac{v^{2}_{\bf k}}{\omega-E_{\bf k}}
\end{array}\right) .
\label{gS}\end{eqnarray} 
In the equation (\ref{gS}) we applied the BCS coefficients 
\begin{eqnarray} 
u^{2}_{\bf k},v^{2}_{\bf k} &=& \frac{1}{2} \left[ 1 \pm 
\frac{\xi_{{\bf k}S}}{E_{\bf k}} \right]
\nonumber \\
u_{\bf k}v_{\bf k} &=& \frac{\Delta}{2E_{\bf k}} ,
\nonumber
\end{eqnarray}
where $E_{\bf k}\!=\!\sqrt{\xi_{{\bf k}S}^{2}+\Delta^{2}}$.

For the case of uncorrelated quantum dots ($U_{i}\!=\!0$) we 
can determine the selfenergy $\Sigma^{0}(\omega)$ 
of the interfacial quantum dot from the following equation
\begin{eqnarray}
\mb{\Sigma}^{0}(\omega) = \!\!\! \sum_{{\bf k},\beta\!=\!N,S} 
V_{{\bf k},\beta} \;\; g^{r}_{\beta}({\bf k},\omega) \; 
V_{{\bf k},\beta}^{*} 
+ t \; g^{r}_{2}(\omega) \; t^{*} .
\label{S_r}
\end{eqnarray}
Assuming the wide-band limit we introduce the constant 
weighed density of states   
\begin{eqnarray}
2\pi\sum|V_{{\bf k} \beta}|^{2}
\delta\left( \omega\!-\!\xi_{{\bf k},\beta}\right)=
\left\{ \begin{array}{l}
\Gamma_{\beta} \hspace{0.2cm} \mbox{\rm for} \hspace{0.2cm} 
|\xi_{{\bf k},\beta}| <D/2\\
0 \hspace{0.2cm} \mbox{\rm elsewhere},
\end{array} \right.
\end{eqnarray}
where $D$ is the conduction bandwidth. We then easily find 
that  
\begin{eqnarray}
\sum_{{\bf k}} V_{{\bf k},N} \;\; g^{r}_{N}({\bf k},\omega) 
\; V_{{\bf k},N}^{*} =
\left( \begin{array}{cc}  
\frac{-i\Gamma_{N}}{2} & 0 \\ 
0 & \frac{-i\Gamma_{N}}{2} 
\end{array} \right)
\label{Sigma_N}
\end{eqnarray} 
because, according to  the Kramers-Kr\"onig relation, the real 
part disappears. In the same way we obtain from a straightforward 
algebra that \cite{review-paper}
\begin{eqnarray}
\sum_{{\bf k}} V_{{\bf k},S} \; g^{r}_{S}({\bf k},\omega) 
\; V_{{\bf k},S}^{*} = \frac{\gamma(\omega)\Gamma_{S}}{2i} 
\left( \begin{array}{cc}  
-1 & \frac{\Delta}{\omega} \\ 
 \frac{\Delta}{\omega}  & -1 
\end{array} \right)
\label{Sigma_S}
\end{eqnarray} 
where
\begin{eqnarray}
\gamma(\omega) = 
\frac{|\omega| \; \Theta(|\omega|\!-\!\Delta)}
{\sqrt{\omega^{2}-\Delta^{2}}}
-\frac{i \omega \; \Theta(\Delta\!-\!|\omega|)}
{\sqrt{\Delta^{2}-\omega^{2}}} .
\label{gamma}
\end{eqnarray} 
In the extreme subgap limit $|\omega| \ll \Delta$ the function 
(\ref{gamma}) approaches $\gamma (\omega) \rightarrow -i 
\omega /\Delta $ and in consequence 
\begin{eqnarray} \lim_{|\omega| \ll \Delta}
\mb{\Sigma}^{0}(\omega) = 
\left( \begin{array}{cc}  
\frac{-i\Gamma_{N}}{2} + \frac{|t|^{2}}{\omega-\varepsilon_{2}} 
& \frac{-\Gamma_{S}}{2} \\ \frac{-\Gamma_{S}}{2} & 
\frac{-i\Gamma_{N}}{2}  + \frac{|t|^{2}}{\omega+\varepsilon_{2}} 
\end{array} \right)
\label{prooved}
\end{eqnarray}
which proves the equation (\ref{Sigma0}).

\end{document}